\documentclass[epj]{svjour}

\usepackage{graphicx}
\usepackage{fancyhdr}
\usepackage{verbatim}
\def\makeheadbox{{%  remove EPJC header
 }}
\def\mytitle{My title} 
\def\myauthors{My name}  
\def\mytype{My type of session}
\def\mysession{My session}
\def\mytitle{Measuring Fine Tuning In Supersymmetry} %Put your title here!
\def\myauthors{Peter Athron}    %Put your name here!
\def\mytype{Contributed Talk}    
\def\mysession{Theoretical Models}

\begin{document}
\title{Measuring Fine Tuning In Supersymmetry}
%%\subtitle{Do you have a subtitle?\\ If so, write it here}
\author{Peter Athron\inst{1}
\thanks{\emph{p.athron@physics.gla.ac.uk}}%
 \and
 D.J.~Miller\inst{1}% etc
%%\thanks{\emph{Present address:} Insert the address here if needed}%
}                     % Do not remove
\institute{Department of Physics and Astronomy, University of Glasgow, Glasgow G12 8QQ, UK}
\date{}
\abstract{The solution to fine tuning is one of the principal motivations for supersymmetry. However constraints on the parameter space of the Minimal Supersymmetric Standard Model (MSSM) suggest it may also require fine tuning (although to a much lesser extent).  To compare this tuning with different extensions of the Standard Model (including other supersymmetric models) it is essential that we have a reliable, quantitative measure of tuning. We review the measures of tuning used in the literature and propose an alternative measure. We apply this measure to several toy models and the MSSM with some intriguing results. % end of PACS codes
} %end of abstract
\maketitle
\section{Introduction}
\label{intro}

The Little Hierarchy Problem arose when no Beyond the Standard Model (BSM) physics was found at LEP, despite expectations from naturalness that it would. In particular Barbieri and Giudice \cite{Barbieri:1987fn} argued that to avoid fine tuning the supersymmetric particles of the Constrained Minimal Supersymmetric Standard Model (CMSSM) should be within the mass reach of LEP. 

The mass of the $Z$ boson is predicted from the soft supersymmetry (susy) breaking parameters by imposing electroweak symmetry breaking conditions.  For \mbox{$\tan{\beta}= 10$}, 

\noindent \mbox{$\frac{M_Z^2}{2}\approx  -|\mu|^2 + 0.076 m_0^2 + 1.97 m_\frac{1}{2}^2 + 0.1 A^2 + 0.38 A m_\frac{1}{2}$}, where the parameters are $m_0$, the universal scalar mass; $m_{1/2}$, the universal gaugino mass;  $A$, the universal trilinear coefficient; ${\rm sign}(\mu)$, the undetermined sign of $\mu$, a bilinear soft Higgs mass, and $\tan \beta$. 
 Although the correct $M_Z = 91.1876$ GeV can be obtained by fixing $|\mu|$, if $1.97 m_{1/2} \approx 500$ then this must cancel with some combination of parameters to ${\cal O}( 1 / 25 )$.

\begin{comment}
The desire to solve this ``Little Hierarchy Problem'' has motivated a
flood of activity in the construction of supersymmetric models
\cite{Kitano:2006gv,Kitano:2005wc,Nomura:2005qg,Lebedev:2005ge,Chang:2006ra,Choi:2005uz,Choi:2005hd}. There
is also increased interest in studying alternative solutions to the SM
Hierarchy problem \cite{Chacko:2005pe,Chacko:2005un,Casas:2005ev}. In
addition to ensuring such models satisfy phenomenological constraints
it is essential that the naturalness is examined using a reliable,
quantitative measure of tuning.  \\
\end{comment}

  To quantify tuning Barbieri and Guidice applied a measure originally proposed in Ref.\cite{Ellis:1986yg}.  For an observable, $O$, and a parameter, $p_i$, \mbox{$\triangle_{BG}(p_i) = \Big{|}\frac{p_i}{O(p_i)}\frac{\partial O(p_i)}{\partial p_i}\Big{|} \rm{.}$}
%

\begin{comment}
\begin{equation}              
\label{Trad_meas}
\triangle_{BG}(p_i) = \Big{|}\frac{p_i}{O(p_i)}\frac{\partial O(p_i)}{\partial p_i}\Big{|} \rm{.}
\end{equation}
%
\end{comment}
A large value of $\triangle_{BG}(p_i)$ implies that a small \linebreak change in
the parameter results in a large change in the observable, so the
parameters must be carefully ``tuned'' to the observed value.  Since
there is one \linebreak $\triangle_{BG}(p_i)$ per parameter, they define the
largest of these values to be the tuning for that scenario, \linebreak $\triangle_{BG} = \rm{max}(\{\triangle_{BG}(p_i)\})\rm{.}$ They then make the aesthetic choice that $\triangle_{BG} > 10$ is fine tuned.

Despite wide use of $\triangle_{BG}$, it has several limitations
which may obscure the true picture of tuning:
\begin{itemize}\setlength{\parskip}{0ex}
\item variations in each parameter are considered separately;
\item only one observable is considered in the tuning measure, but there may be tunings in several observables;
\item only infinitesimal variations in the parameters are considered;
\item there is an implicit assumption that the parameters come from uniform probability distributions.
\item it does not take account of global sensitivity;
\end{itemize}
%
\begin{comment}
\noindent Tuning is really concerned with how the parameters combine
to produce an unnatural result. So a tuning measure should vary all of
the parameters simultaneously. Secondly, some theories may contain
significant tunings in more than one observable. We also want to know
how can these tunings be combined to provide a single measure.
 
Another concern is that $\triangle_{BG}$ only considers infinitesimal
variations in the parameters. Since MSSM observables are complicated
functions of many parameters, there may be locations where some
observables are stable (unstable) locally, but unstable (stable) over
finite variations.

There is also an implicit assumption that all values of the parameters
in the effective softly broken Lagrangian ${\cal L}_{SUSY}$ are
equally likely. However they have been written down in ignorance
of the high-scale theory, and may not match the parameters in
the high-scale Lagrangian, e.g.~${\cal L}_{GUT}$. Any non-trivial
relation between these different sets of parameters may alleviate or
exacerbate the fine tuning problem.
\end{comment}

%%\subsection{Global Sensitivity}
%%\label{Global Sensitivity}

\noindent The final problem can be understood by considering the
simple mapping $f: x \rightarrow x^n $, where $n \gg 1$. For this
function $\triangle_{BG} = \triangle_{BG}(x) = n$. Since
$\triangle_{BG}$ is independent of $x$, we follow the example of
\cite{Anderson:1994dz} and term this {\em global sensitivity}. Since
$\triangle_{BG}(x_1) - \triangle_{BG}(x_2) = 0$ for all $x_1, x_2$,
there is no {\em relative sensitivity} between points in the parameter
space.

If we use $\triangle_{BG}$ as our tuning measure then $f(x)$ appears
fine tuned throughout the entire parameter space. This contrasts with
our fundamental notion of tuning being a measure of how atypical a
scenario is.  A true measure of tuning should only be greater than
one when there is relative sensitivity between different points in
the parameter space.

\section{A New Tuning Measure}
\label{A_New_Measure}

We propose a new measure of tuning. We define two volumes in parameter
space for every point $P' \{p_i'\}$.  $F$ is the volume formed from
dimensionless variations in the parameters over some arbitrary range
$[a,b]$, about point $P'$, i.e.~\mbox{$a\leq\frac{p_i}{p_i'}\leq b$}.
$G$ is the volume in which dimensionless variations of the observables
fall into the same range $[a,b]$,
i.e. $a\leq\frac{O_j(\{p_i\})}{O_j(\{p_i'\})}\leq b$.

We define an unnormalised measure of tuning with, $\triangle = \frac{F}{G} \rm{.}$
This can be used to compare different regions of parameter space within a given model as the normalisation factor will be common.  However like $\triangle_{BG}$ it includes global sensitivity. To compare tuning in different models we need to include normalisation, so tuning is given by,
\begin{equation} 
 \triangle = \frac{F}{G} \rm{,}\,\,\,\,\;\;\;\;
\hat{\triangle} = \frac{1}{\bar{\triangle}}\frac{F}{G} \rm{,}
\end{equation}
\begin{equation} 
\rm{with},\, \bar{\triangle} = \left\langle\frac{F}{G}\right\rangle 
= \frac{\int dp_1 ... dp_n 
\frac{F}{G}(\{p_i\}, \{O_i\})}{\int dp_1 ... dp_n}\rm{.}
\end{equation}

It is also useful to look at tuning in terms of individual observables, while maintaning our multi-parameter approach.
  Therefore we define $G_{O_j}$ to be the
volume restricted by $a\leq\frac{O_j(\{p_i\})}{O_j(\{p_i'\})}\leq b$
and $a\leq\frac{p_i}{p_i'}\leq b$. Tuning is then defined by,
\begin{equation} 
\hat{\triangle}_{O_j} 
= \frac{1}{\left\langle\frac{F}{G_{O_j}}\right\rangle}\frac{F}{G_{O_j}} \rm{,}
\end{equation}
%
%%This definition is applied to obtain individual tunings in the MSSM in Section \ref{MSSM_Tuning}. 

\subsection{Probabilistic Interpretation}

$G$ is a volume containing physical scenarios which are ``similar'' to point P, where our notion of similarity is given by $a$, $b$. Assuming every point in parameter space is equally likely, the probability of a randomly selected point lying in $G$ is $G/V_T$, where $V_T$ is some hypothesised total volume of parameter space.  If there existed some volume, $T$,  which was ``typical'' in size, the probability of a random point lying in it would be  $T/V_T$.  Our expectation for the volume $G$ is based on the magnitude of the parameters.  The volume formed by ``similar'' parameters is a way of combing the magnitudes of each parameter into one one volume which describes them all.  This is our volume $F$.  Knowing only the volume of $F$ we can find the $G$ we would typically expect by comparing it to the average ratio between $F$ and $G$. %%Then  $T = F/\bar{\triangle}$ gives us our typical expectation for the volume for $G$, given the size of $F$.
So for parameters ``similar'' in size to those at point P, the volume one typically expects ``similar'' physical scenarios to form  is $T = F/\bar{\triangle}$. Therefore we can associate our new measure of tuning with relative improbability, $\hat{\triangle} \sim \frac{P(T)}{P(G)}$.

\section{Applications}

  As a first example we determine the tuning for a toy version of the Standard Model (SM) Hierarchy Problem with only one observable, the physical Higgs \linebreak (mass)$^2$, $m_H^2$, and one parameter, $m_0^2$. At one loop we
write, $m^2_H = m^2_0 - C\Lambda^2$, where $\Lambda$, the Ultra-Violet cutoff, is taken to be the Planck Mass, while  $C$ is a positive constant.

 Variations in $m_0^2$ give a line of length $F=(b-a)m_0^2$, while variations in the observable, $m_H^2$, give another line, of length $G = (b-a)m_H^2$.
\begin{equation}\Rightarrow \triangle = m_0^2/m_h^2 = \triangle_{BG} \approx 10^{34}\end{equation} The arbitrary range $[a,b]$ has fallen out of the result.

We also determine $\bar{\triangle}$, by integrating over the whole parameter range, $m^2_{0\, min} < m_0^2 < m_{0\,max}^2$ where $m^2_{0\, max}$, $m^2_{0\, min}$ are hypothetical upper and lower limits respectively and present results where $m_{H}^2 > 0$. These bounds give the
total allowed range of the parameter in this model and should not be
confused with the range of dimensionless variations which appears in
the definition of $F$. If we take the range of variation to be large,
\mbox{$m_{0 max}^2 - m_{0 min}^2 \gg C \Lambda^2$}, then \mbox{$
\hat{\triangle} \approx \frac{m_0^2}{m_H^2} = \triangle_{BG}$}.
Alternatively, if we choose a very narrow range of variation about $C
\Lambda^2 + \mu_H^2$, where $\mu_H \approx 100 \,\rm{GeV}$, then
$\hat{\triangle}$ is very small. 

This is intuitively reasonable.  If there was some compelling
theoretical reason for the bare mass to be constrained close to the
cutoff, the case for new physics at low energies would be dramatically
weakened.  It is precisely because there is no such compelling reason
that we worry about the hierarchy problem and look to low energy
BSM physics to explain how we can have $m_H \ll M_{\rm Planck}$.

Now lets treat the SM Hierarchy Problem in a slightly more sophisticated way . We no longer fix the UV cutoff.  Instead we treat $m_H^2$ as a function of two parameters, $m_0^2$ and $\Lambda^2$ and we have a second observable
$M^2_{\rm Planck} = \Lambda^2 $ (``observed'' to be large due to the weakness of gravitation).
%
\begin{comment}\begin{equation} 
M_{\rm Planck}^2 
=  \Lambda^2,\: \: \: \:\quad m^2_H = m^2_0 - C\Lambda^2 .
\end{equation}     
\end{comment}

 There are no new cancellations between the parameters, so we expect
the same result for $\triangle$ as before, but this provides a simple
illustration of how our measure works with more than one
parameter. Varying the parameters about some point $P'(m_0^2,
\Lambda^2)$ over the dimensionless interval $[a,b]$ forms an area,
$F$. The bounds from dimensionless variations in $m_H^2$ introduce two
new lines in the parameter space which together with dimensionless
variations in $M_{\rm Planck}^2$ (the same as those from $\Lambda^2$),
form the area $G$.

\begin{figure}[ht]
\begin{center}
\includegraphics[height=50mm]{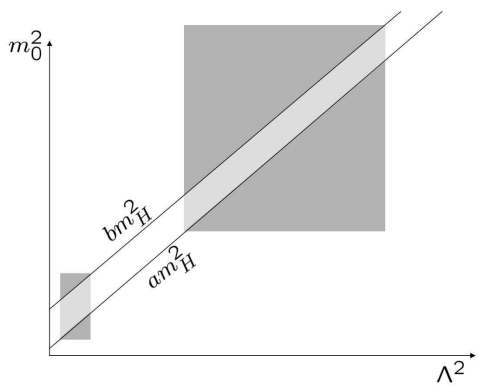}
\end{center}
\caption{The two dimensional volumes (areas) $F$ (entire grey
rectangle) and $G$ (light grey) for two different points in the two
dimensional parameter space.}
\label{2d_SM}
\end{figure}

 This is shown schematically in Fig.~\ref{2d_SM} for two different points. In one,
the parameters are of the same order as the observables (since $M_{
\rm Planck}$ is {\cal O}($m_H$)), so $G$ is not much smaller than $F$.
For the other point $M^2_{\rm Planck} \gg m_H^2$, resulting in  $F \gg G$ and fine tuning. In general the areas are, $F =
(b-a)^2m_0^2\Lambda^2$ and $G = (b-a)^2\Lambda^2m_H^2$ so again we
obtain, $\triangle = 1 + \frac{C\Lambda^2}{m_H^2} = \triangle_{BG}$.

 While our measure does not deviate from
$\triangle_{BG}$ in this simple example, models with additional
parameters allow the observable to be obtained from cancellation of
more than two terms, complicating the fine tuning picture.   For more examples including one with three parameters and four observables please see Ref.~\cite{Athron:2007ry}.

\subsection{Fine Tuning In the CMSSM}

 Since the CMSSM contains many parameters and many observables we chose to apply a numerical version of our measure to study the variation of tuning in the CMSSM. 

 We take random dimensionless fluctuations about a CMSSM point at the
 GUT scale, $P' = \{p_k\}$, to give new points $\{P_i\}$. These are
 passed to a modified version of Softsusy
 2.0.5\cite{Allanach:2001kg}. Each random point $P_i$ is run down from
 the GUT scale until electroweak symmetry is broken. An iterative
 procedure is used to predict $M_Z^2$ and then all the sparticle and
 Higgs masses are determined.
\begin{comment}
As before $F$ is the volume formed by dimensionless variations in the
parameters. $G_{O_i}$ is the sub-volume of $F$ additionally restricted
by dimensionless variations in the single observable $O_i$,
$a\leq\frac{O_i(\{p_k\})}{O_i(\{p_k'\})}\leq b$. As usual $G$ is the
volume restricted by $a\leq\frac{O_j(\{p_k\})}{O_j(\{p_k'\})}\leq b$,
for each observable, $O_j$ where $\{O_j\}$ is the set of masses
predicted in Softsusy.
\end{comment}

 For every observable $O_i$ a count, $N_{O_i}$, is kept of
how often the point lies in the volume $G_{O_i}$ as well as an overall
count, $N_O$, kept of how many points are in $G$. The tunings are then
measured with,
\begin{equation}\triangle_{O_i} \approx
\frac{N}{N_{O_i}}{\rm ,}\,\,\,\triangle \approx \frac{N}{N_O}.\end{equation}

 The set of observables, $\{O_j\}$ used in our definition of $G$ here is the set of $M_Z^2$ and all (masses)$^2$ predicted in Softsusy.

The parameters we vary simultaneously are the set\footnote{Note that
all CMSSM points have $|\mu|$ set by $M_Z^2$, so our tuning measure is
not sensitive to the $\mu$-problem.  However for our random variations
we do treat $\mu_{GUT}$ as a parameter because we are predicting
$M_Z^2$, not fixing it to it's observed value. } $\{m_0, m_{1/2},
\mu_{GUT}, m_3^2, A, y_t, y_b, y_{\tau} \}$, where $m_3$ is the soft
bilinear Higgs mixing parameter and $y_t, y_b, y_{\tau}$ are the
Yukawa couplings of the top, bottom and tau respectively

When using Softsusy to predict the masses for the random
points, sometimes the full mass spectrum cannot be predicted as we may have a tachyon, the Higgs potential unbounded from below, or non-perturbativity. Such
points don't belong in $G$ as they will give dramatically
different physics. However it is unclear which volumes, $G_{O_i}$, the
point lies in. Such points never register as hits in any of the
$G_{O_i}$ and this may artificially inflate the individual tunings,
including $\triangle_{M_Z^2}$. Keeping the range small reduces such errors, so we chose $a=0.9$ and $b=1.1$ for our dimensionless variations.
\begin{comment}
Also, since we are measuring tuning for individual points
numerically and cover only a small sample of points, it is not
possible to obtain mean values of $\triangle$ and the
$\triangle_{O_i}$ as we haven't sampled the entire space. When
simply comparing how the tuning varies about the parameter space the
normalisation factor is not needed. However to compare the tuning
between different observables as well as to compare with different
models some form of normalisation is essential.
\end{comment}
 %%The gauge couplings are not included as parameters. Doing so would introduce excessive global sensitivity, increasing the statistics needed to keep the errors under control.

 We examine tuning for points in the grid,
\begin{eqnarray*}A = -100 \,{\rm GeV},\;\;\;\; 
\tan{\beta}= 10,\;\;\; {\rm sign}(\mu) = +, \nonumber \\  
250\,{\rm GeV} \leq m_{\frac{1}{2}} \leq 500\,{\rm GeV}{\rm ,} \; 
100\,{\rm GeV} \leq m_{0} \leq 200\,{\rm GeV}.
\end{eqnarray*}

\begin{figure}[h!]
\begin{center}

\includegraphics[height=55mm,clip=true]{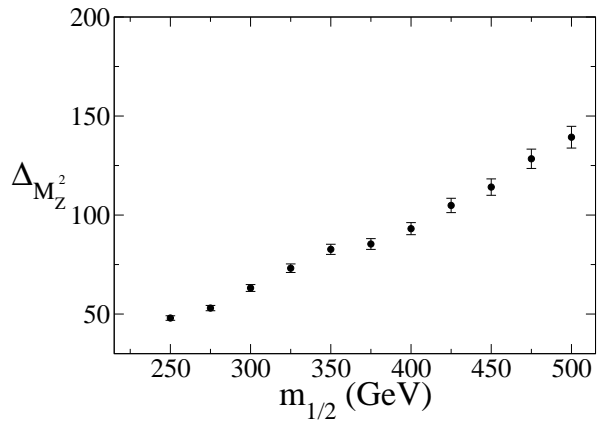} \hfill

\end{center}
\caption{Tuning variation in $M_Z^2$
plotted against $m_{1/2}$.  To reduce statistical errors, at each
value of $m_{1/2}$, we have taken the mean value $\triangle_{M_Z^2}$
over the five different $m_0$ values. }
\label{Tuning_Mz_fig}
\end{figure}

Shown in Fig.\ref{Tuning_Mz_fig} is the variation in $\triangle_{M_Z^2}$ with
respect to $m_{1/2}$.  To reduce statistical errors the
$\triangle_{M_Z^2}$ for each $m_{1/2}$ is averaged over the five
different $m_0$ values. This substantially reduces the errors giving a
much more stable picture of tuning increasing linearly with $m_{1/2}$.

\begin{figure}[ht]
\begin{center}
\includegraphics[height=55mm,clip=true]{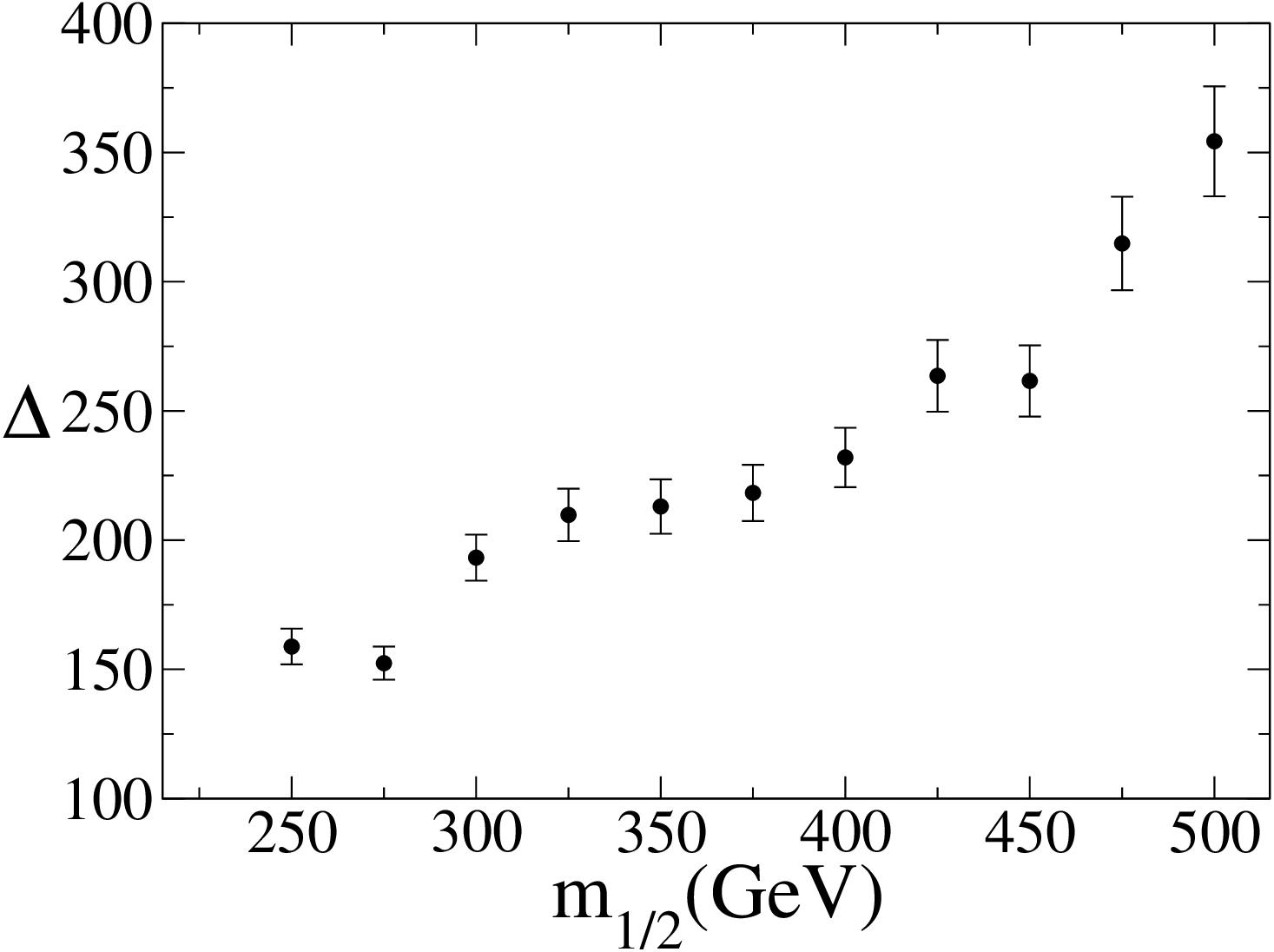} \hfill
\end{center}
\caption{Variation in $\triangle$ plotted as in
Fig.~\ref{Tuning_Mz_fig} for $\triangle_{M_Z^2}$.}
\label{Tuning_all_figs}
\end{figure}

$\triangle$, which includes all of the masses predicted by Softsusy as
well as $M_Z^2$, is shown in Fig.~\ref{Tuning_all_figs}. Although the
errors are much larger here, a similar pattern to that for $M_Z^2$ can
be seen.  Since these are unnormalised tunings, the numerical values
of the two measures cannot be compared and one should not assume that
$\triangle > \triangle_{M_Z^2}$ implies that the tuning is worse than
when only $M_Z^2$ was considered.  In fact the lack of evidence for
distinct patterns of variation in tuning from the
Figs.~\ref{Tuning_Mz_fig} and \ref{Tuning_all_figs} is consistent with
the conjecture that the large cancellation between parameters in
$M_Z^2$ is the dominant source of the tuning for these points.

Although we can't easily determine the normalisation using this approach it
is nonetheless interesting to compare the unnormalised tunings for the
points in our study with those obtained for points with more
``natural'' looking spectra.  We present two points for this
purpose. NP1 and NP2 are defined by,   

\noindent$\{m_{\frac{1}{2}} = M_Z {\rm ,}\,\, m_0= M_Z,\,\, A =  -M_Z, \,\,{\rm sign}(\mu) = +,\\ \tan \beta = 3\}$ and 
\noindent$\{m_{\frac{1}{2}} = -50\, \rm{GeV} {\rm ,}\,\, m_0= 100\, \rm{GeV}, \\ A =  -50\, \rm{GeV}, \,\,{\rm sign}(\mu) = +, \tan \beta = 10\}$ respectively.
%%\noindent\mbox{$ m_{\frac{1}{2}} = -50 \rm{GeV} {\rm ,} m_0= 100 \rm{GeV}, A =  -50 \rm{GeV}, {\rm sign}(\mu) = +, \tan \beta = 10$} respectively.
%%\begin{eqnarray*} m_{\frac{1}{2}} = -50 \rm{GeV} {\rm ,} m_0= 100 \rm{GeV}, A =  -50 \rm{GeV}, \\ {\rm sign}(\mu) = +, \tan \beta = 10  \end{eqnarray*}

\begin{figure}[ht]
\begin{center}
\includegraphics[height=40mm, clip=true]{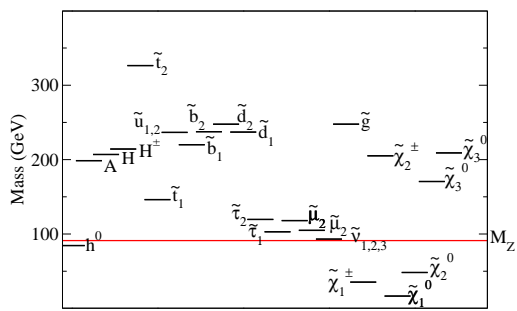}
\end{center}
\caption{Point NP1 with a ``natural'' spectrum}
\label{NP_1_spectrum}
\end{figure}

\begin{figure}[ht]
\begin{center}
\includegraphics[height=40mm, clip=true]{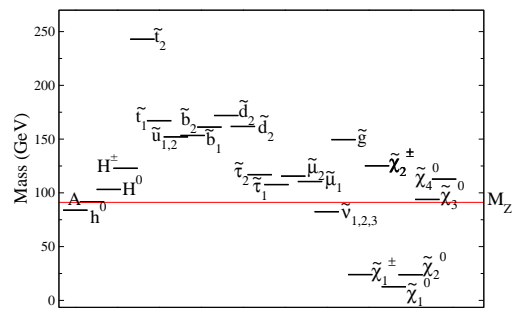}
\end{center}
\caption{Point NP2 with a ``natural'' spectrum}
\label{NP_2_spectrum}
\end{figure}

\begin{comment}
\begin{equation}
 \begin{array}{lrclrclrclrclrcl}
 \hspace*{-3mm} {\rm NP1:} 
 & m_{\frac{1}{2}} & \hspace*{-3mm} = \hspace*{-3mm} & M_Z{\rm ,} 
& m_0 &\hspace*{-3mm} = \hspace*{-3mm}& M_Z{\rm ,}
& a_0 &\hspace*{-3mm} = \hspace*{-3mm}& -M_Z{\rm ,}
& {\rm sign}(\mu) &\hspace*{-3mm} = \hspace*{-3mm}& +{\rm ,}
& \tan{\beta} &\hspace*{-3mm} = \hspace*{-3mm}& 3, \\
\hspace*{-3mm} {\rm NP2:} 
& m_{\frac{1}{2}} & \hspace*{-3mm} = \hspace*{-3mm} & -50\, {\rm GeV,} 
& m_0 &\hspace*{-3mm} = \hspace*{-3mm}& 100\, {\rm GeV,} 
& a_0 &\hspace*{-3mm} = \hspace*{-3mm}& -50\, {\rm GeV,} 
& {\rm sign}(\mu) &\hspace*{-3mm} = \hspace*{-3mm}& +{\rm ,}
& \tan{\beta} &\hspace*{-3mm} = \hspace*{-3mm}& 10.
\end{array}
\end{equation}
%
\end{comment}
 The spectra of these points are displayed in
Fig.~\ref{NP_1_spectrum} and Fig.~\ref{NP_2_spectrum}, and the
unnormalised tunings are displayed in Table
\ref{Tuning_comp_points}. Note that these are not intended to be
``realistic'' scenarios. Indeed both NP1 and NP2 are ruled out by
experiment but are simply intended to provide ``natural'' scenarios
for comparison.

\setlength{\tabcolsep}{4mm}
\begin{table}
\begin{center}
\begin{tabular}{l|cccccc} 
& $\triangle$ 
& $\triangle_{M_Z^2}$   
& $\triangle_{m_{\chi_1^0}^2}$\\ \hline\\
{\rm NP1}  
& $241^{+36}_{-26}$ 
& $14.7^{+0.5}_{-0.5}$   
& $30.1^{+1.4}_{-1.3}$\\ \\
{\rm NP2}       
& $31.4^{+1.5}_{-1.4}$  
& $2.92^{+0.04}_{-0.04}$  
& $2.64^{+0.04}_{-0.04}$\\  
\end{tabular}
\caption{Unnormalised tunings for the two points, NP1 and NP2, with
natural looking spectra.}
\label{Tuning_comp_points}
\end{center}
\end{table}

While in  NP1 $\triangle_{M_Z^2}$ is reduced there's a
relatively large tuning in the mass of the lightest neutralino
($\triangle_{m_{\chi_1^0}^2}$). These combine to give a $\triangle$
which is similar in size to the values found for our grid of
points. In NP2 all of the tunings are relatively small, but the
combined tuning is still larger than may naively have been
anticipated. This is because many of these small tunings for individual
observables, restricting different regions
of parameter space. Table \ref{Relative_Tunings} shows the approximate
relative magnitude of the tunings in our grid points with respect to
these seemingly natural points.

\begin{table}
\begin{center}
\begin{tabular}{l|cccccc} 
& $\hat{\triangle}$ 
& $\hat{\triangle}_{M_Z^2}$   
& $\hat{\triangle}_{m_{\chi_1^0}^2}$\\ \hline\\
{\rm Relative to NP1}\,   
& $0.5-1.5$ 
& $3-10$ 
& $0.2 $\\ \\
{\rm Relative to NP2}\,   
& $5-15 $  
& $10-50$  
& $2$\\  
\end{tabular}
\caption{Approximate relative tunings for the points in our study,
with respect to those for NP1 and NP2.}
\label{Relative_Tunings}
\end{center}
\end{table}

Notice that if for either NP1 or NP2 we truely had $\triangle^{NP} \approx \bar{\triangle}$ and $\triangle_{M_Z}^{NP} \approx \bar{\triangle}_{M_Z} $ then we would have demonstrated that $\hat{\triangle}_{M_Z} > \hat{\triangle}$ and could conclude that the Little Hierarchy problem is not as severe as has been suggested.   

 Sometimes (e.g.~NP1) the lightest neutralino
is very light due to large cancellations between the
parameters. Similar effects may be present in other masses, so
mass hierarchies may appear in a greater proportion of the parameter space
than conventional CMSSM wisdom dictates.  This would reduce the true
tuning in the CMSSM as scenarios with hierarchies would be less
atypical than previously thought.  A reduction in tuning from this
effect can only be measured by using our normalised new measure,
$\hat{\triangle}$.

\begin{comment}
Unfortunately the numerical approach we have applied to the MSSM in
this paper cannot be used to address this issue. An average measure of
$\triangle$, over the whole parameter space, is needed in order to
investigate this possibility.  A thorough numerical survey of the
parameter space would be too expensive, however an analytical study
may be more promising.  Findings in numerical studies like this may be
used to identify which observables and parameters are important for
fine tuning and therefore reduce the set $\{O_i\}$ and $\{p_i\}$ to a
manageable size. We will not carry out this programme here, but leave
it for a future study.
\end{comment}

\section{Conclusions}

Current measures of tuning have several limitations. They neglect the
many parameter nature of fine tuning; ignore additional tunings in
other observables; consider local stability only; assume ${\cal L}_{SUSY}$ is pa\-ram\-e\-trised in the same way as ${\cal L}_{GUT}$ and do not account for global sensitivity. 
 
We have presented a new measure of tuning to address these issues.  We showed that in the CMSSM both $\triangle$ and $\triangle_{BG}$ increase with $M_{1/2}$.  While a naive interpretation suggests $\triangle_{BG} > \triangle$ normalisation may dramatically change this. If $\hat{\triangle} <<  \hat{\triangle}_{M_Z}$ we can explain the Little Hierarchy Problem.


\begin{thebibliography}{999}

%\cite{Barbieri:1987fn}
\bibitem{Barbieri:1987fn}
  R.~Barbieri and G.~F.~Giudice,
  %``Upper Bounds On Supersymmetric Particle Masses,''
  Nucl.\ Phys.\  B {\bf 306}, 63 (1988).
  %%CITATION = NUPHA,B306,63;%%

\bibitem{Ellis:1986yg}
  J.~R.~Ellis, K.~Enqvist, D.~V.~Nanopoulos and F.~Zwirner,
  %``Observables In Low-Energy Superstring Models,''
  Mod.\ Phys.\ Lett.\  A {\bf 1} (1986) 57.
  %%CITATION = MPLAE,A1,57;%%


%\cite{Anderson:1994dz}
\bibitem{Anderson:1994dz}
  G.~W.~Anderson and D.~J.~Castano,
  %``Measures of fine tuning,''
  Phys.\ Lett.\  B {\bf 347}, 300 (1995)
  [arXiv:hep-ph/9409419].
  %%CITATION = PHLTA,B347,300;%%




%\cite{Allanach:2001kg}
\bibitem{Allanach:2001kg}
  B.~C.~Allanach,
  %``SOFTSUSY: A C++ program for calculating supersymmetric spectra,''
  Comput.\ Phys.\ Commun.\  {\bf 143}, 305 (2002)
  [arXiv:hep-ph/0104145].
  %%CITATION = CPHCB,143,305;%%

\bibitem{Athron:2007ry}
  P.~Athron and D.~J.~Miller,
  %``A New Measure of Fine Tuning,''
  arXiv:0705.2241 [hep-ph].
  %%CITATION = ARXIV:0705.2241;%%



\end{thebibliography}
\end{document}